# Interpretable discovery of new semiconductors with machine learning

Hitarth Choubisa[*,1], Petar Todorović[*,1], Joao M. Pina[1], Darshan H. Parmar[1], Ziliang Li[1], Oleksandr Voznyy[4], Isaac Tamblyn[†,2,3], Edward Sargent[†,1]

[1]Department of Electrical and Computer Engineering, University of Toronto, 10 King's College Road, Toronto, ON, M5S 3G4, Canada.

[2]National Research Council of Canada, Ottawa, ON, K1A 0R6, Canada.

[3]Vector Institute for Artificial Intelligence, Toronto, ON, M5G 1M1, Canada.

[4]Department of Physical and Environmental Sciences, University of Toronto, Scarborough, ON, M1C 1A4, Canada.

[*] These authors contributed equally to this work.

[(†)] Correspondence and requests for materials should be addressed to Edward H. Sargent (ted.sargent@utoronto.ca) and Isaac Tamblyn (isaac.tamblyn@nrc.ca)

**Abstract**

Machine learning models of materials[1–5] accelerate discovery compared to *ab initio* methods: deep learning models now reproduce density functional theory (DFT)-calculated results at one hundred thousandths of the cost of DFT[6]. To provide guidance in experimental materials synthesis, these need to be coupled with an accurate yet effective search algorithm and training data consistent with experimental observations. Here we report an evolutionary algorithm powered search which uses machine-learned surrogate models trained on high-throughput hybrid functional DFT data benchmarked against experimental bandgaps: Deep Adaptive Regressive Weighted Intelligent Network (DARWIN). The strategy enables efficient search over the materials space of ~$10^8$ ternaries and $10^{11}$ quaternaries[7] for candidates with target properties. It provides interpretable design rules, such as our finding that the difference in the electronegativity between the halide and B-site cation being a strong predictor of ternary structural stability. As an example, when we seek UV emission, DARWIN predicts $K_2CuX_3$ (X = Cl, Br) as a promising materials family, based on its electronegativity difference. We synthesized and found these materials to be stable, direct bandgap UV emitters. The approach also allows knowledge distillation for use by humans.

**Introduction**

Inverse materials design, i.e., prediction of a structure and composition with targeted properties, is used to accelerate materials discovery for light emission, sensing, lasing, energy harvesting and energy storage. Recently, deep learning models have advanced in predicting the materials properties of molecular and inorganic crystal structures[8–10]. However, even with a deep learning acceleration of $10^5$, such brute-force exploration of the compositional and structural space remains infeasible: there are ~$10^8$ inorganic ternary, ~$10^{11}$ quaternary compounds, and even more variations for alloyed and multinary compositions. An efficient and interpretable search algorithm is required to complement accurate property evaluation.

This combinatorial space may have multiple local optima, hindering the efficiency of gradient-based methods. Searching within a lower-dimensional latent space has been suggested previously[11,12], yet the continuous nature of the latent space – distinct from the discrete nature of chemical elements – leads to multiple predictions based on artificial elements, which when discretized to true elements do not present the optimal solutions due to the sampling methods involved in prediction.

Evolutionary algorithms (EA) are a gradient-free bias-free approach that allow for the rapid exploration of a space[13,14]. They are particularly well-suited to search when the evaluation cost per instance is small compared to the size of the space[15,16]. The evolutionary algorithm approach assesses an entire population at each specific time-step, giving rise to multiple local solutions while iterating towards the global minima[17]. Furthermore, the approach is highly flexible, allowing for the multi-objective optimization of mixed variables (discrete and continuous) thereby enabling efficient space exploration in which typical gradient-based approaches suffer from[18,19].

**Discussion**

Several studies have discussed machine learning (ML) algorithms for accelerated discovery of novel materials, but a major hurdle in actually discovering a new material is the availability of high-fidelity training data. All the open-source material databases[20–22] report Perdew-Burke-Ernzerhof (PBE)[23] exchange-correlation based calculated Density Functional Theory (DFT) bandgaps which are known to severely and inconsistently underestimate[24] the bandgaps of the materials. ML models trained on such data will yield error-prone and unreliable predictions unable to guide the experiments. This has made most studies[6,25] on accelerated materials discovery focus only on the stability of the materials, ignoring the optical properties (bandgap and nature of bandgaps). Using a generated high-throughput hybrid exchange-correlation based bandgaps database (Supplementary Figures 1-3), we bridge this gap between theory and experiments (Figure 1 where we compare ML predictions with experimental bandgap values). We design a machine-learned framework DARWIN: Deep Adaptive Regressive Weighted Intelligent Network. It enables efficient search over a large chemical space using our ML-based models and proposes a diverse set of candidates with specific target properties.

We develop a graph convolutional neural network (GCNN) architecture[6,10] that incorporates a unique edge attribute, which enables efficient learning on a smaller dataset. This specific reciprocal distance edge attribute helps capture the strength of the relative atomic properties for the varied elements in the crystal system. GCNN generates a global representation of the crystal structure from the chemical features representative of each element at a given node and edge feature that scales as the reciprocal distance between the atoms to reproduce the predictions of accurate *ab initio* calculations (Figure 1a, see Methods for details). Using data from the Materials Project database[14,26], we trained a total energy regressor and a classifier for predicting the direct/indirect nature of bandgaps (Supplementary Figures 4-8 for loss and classification curves). We train our bandgap regressor (Supplementary Figure 9, 10) on a newly generated dataset of 1800 samples (which we release as part of this paper) calculated using HSE functional, known to correlate well with experimental values.

We report an original performance comparison between ML predictions and experimental values of the bandgaps (Figure 1b). Despite training on DFT values, ML predicted bandgaps show good agreement with measured values on a test set of experimentally confirmed compounds. The trained and optimized GCNN models predict HSE06 bandgaps with mean absolute errors (MAE) of 0.53 eV on test data, 0.45 eV on validation data, and energies above hull with MAE of 0.06 eV/atom (Figure 1b, c). Furthermore, we build an accurate classifier for predicting the direct-indirect nature of the materials and obtain an $F_1$-score of 0.8 (Figure 1d). This value is close to previously reported (0.89) study on the direct-indirect classification that was limited only to the Kesterite family of compounds[27]. Using GCNNs allowed going beyond a single crystal family and developing a dependable and accurate classifier for all 7 crystal systems. (Figure 1e for a comparison between ML predictions and DFT).

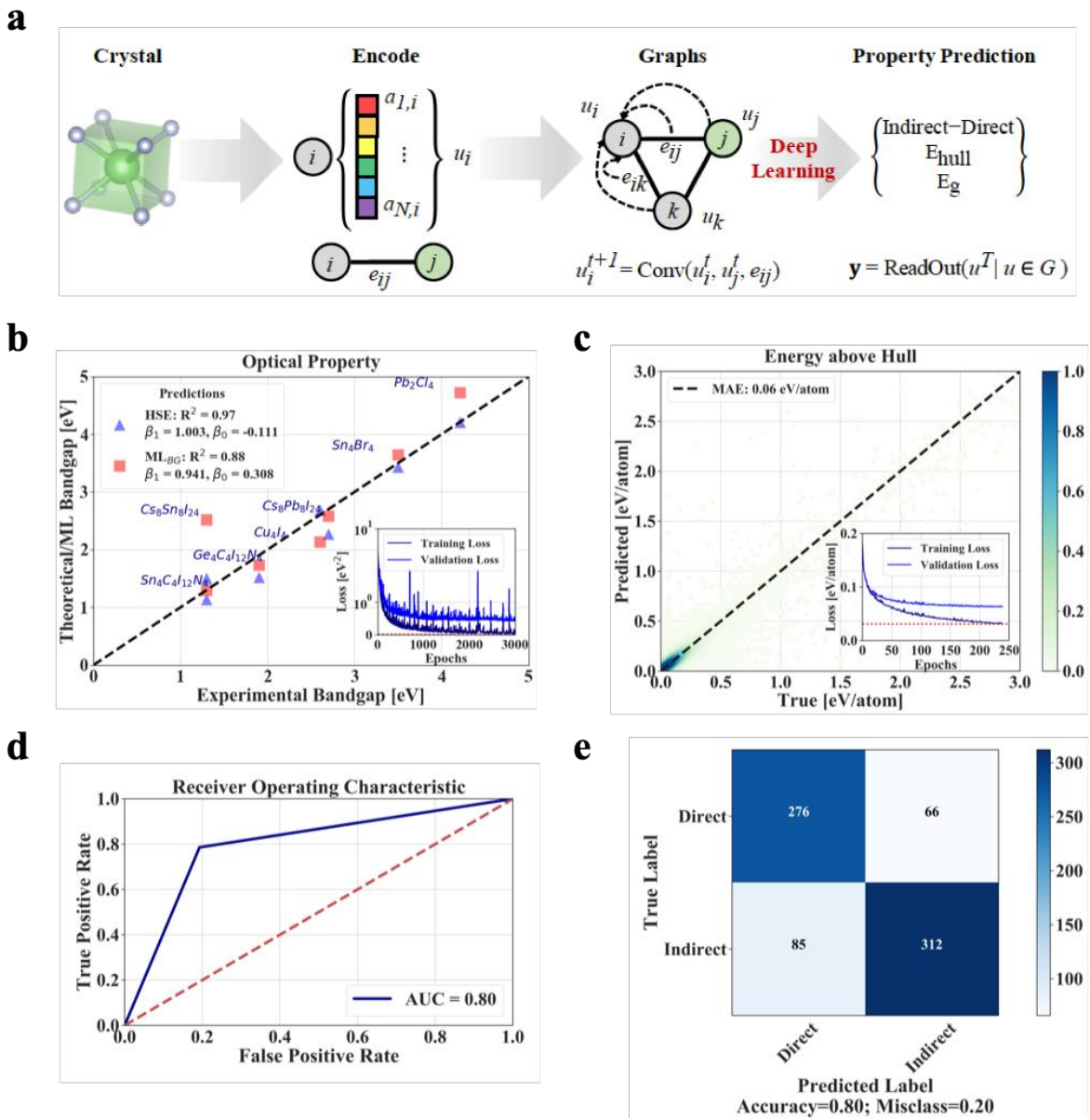

**Figure 1 | Predictive deep learning. a.** Mapping crystals to graph representations through encoding, we train Graph Neural Networks to predict the desired property – bandgaps, energies, and direct/indirect nature; **b.** Performance of the ML model and HSE06 exchange-correlation functional calculated values in predicting experimental bandgaps (inset of training/validation loss curves). The legend includes the coefficient of determination and linear regression parameters as obtained by fitting a linear regression model between the true experimental value and the two

various predictions methods (DFT vs. ML); **c.** Performance of our energy model in predicting total energies which we then use for calculation of energy above hull to determine material stability (inset of training/validation loss curves); **d.** ROC-AUC curve for the direct-indirect bandgap classifier; **e.** Performance of the classifier on direct-indirect classification task.

We then designed a learning-based (i.e. intelligent)[28,29] agent (evolutionary algorithm) to search through materials space (Figure 2) using the optimized GCNNs as surrogate models. The process is divided into (1) random generation of the initial candidates from all possible crystal structures, using compositions that satisfy the charge balance, and (2) the iterative update-evaluation step (Methods) in which candidates are evaluated for fitness and ranked relative to one another based on their fitness score. We set the fitness function as the sum of the mean squared errors of predicted bandgap, energy above hull, and direct-indirect nature against their desired values (Methods) and then ranked accordingly. The bottom half is discarded, and the top half is replicated but with each corresponding structure receiving a mutation, generating a new group of candidates. In our case, we substitute a random element of equivalent oxidation state from a set dictionary of elements available to maintain the charge neutrality of the structure. This procedure results in a new generation of potential solutions of the same initial size.

Evolutionary search starts with a set of randomly generated prototype structures spanning 220 space groups and 7 crystal systems synonymous with the original dataset. The search algorithm relies on the ML surrogate models to predict the properties of interest and evaluate the set of candidates for their fit (Figure 2, and Methods). We found that mutations alone were enough to direct the search toward the optimal compositions of the large chemical space allowing us to skip crossovers as shown by the decreasing loss as generations of solutions proceeds in time (Supplementary Figure 11). Crossover methods potentially proceed towards local-optima and a suboptimal solution as a result of the similarity among the top-ranked individuals – inspiring our single mutation strategy[30]. Although evolutionary search combined with a surrogate model can lead to promising compositions, it does not on its own provide intuitive understanding for the discovery of such materials. The last component of DARWIN enables interpretation[31,32].

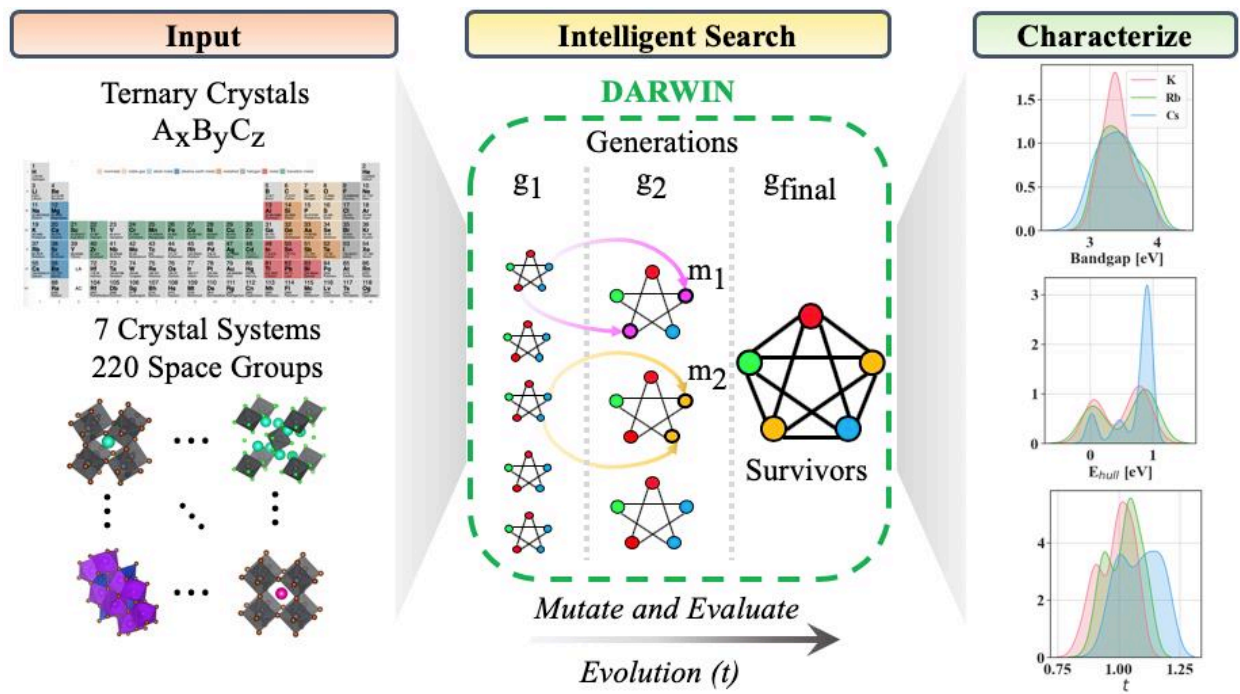

**Figure 2 | Evolutionary Search via surrogate models - DARWIN. Input.** To enable search of compounds with non-rare elements, we limited DARWIN search by excluding Lanthanides, Actinides and rare transition elements. Crystals were generated using substitutions in prototype structures and spanned over 7 crystal systems and 220 space groups; **Intelligent Search.** DARWIN uses trained Graph Networks as surrogate models and mutations to find new candidates; **Characterize.** DARWIN enables the discovery of new compounds and uncovers new chemical trends via unsupervised learning and feature-based ML methods.

We demonstrate the applicability of DARWIN to discover stable and direct UV bandgap materials (3eV - 4eV), a relatively unexplored[28,42] and vast chemical space[7,16]. To expand the list of promising candidates, the search is executed several times using different random initializations. The initialization provides a unique set of optimized crystal structures for each iteration, which we aggregate together to form our predicted dataset (Supplementary Figure 12). We observed that structures of the form: $ABX_4$, $ABX_3$ and $A_2BX_3$ typically exhibited the lowest predicted energy above hull indicating high stability. Top ranking stable materials include examples such as: $Cs_2SnBr_4$, $Rb_2SnBr_4$, $Rb_2CrBr_4$, $CsScBr_4$, $KInBr_4$, and $K_2CuCl_3$ which illustrates elemental diversification among the predictions.

DARWIN yields several promising UV structures from which we select ternary candidates with A-site containing one of the common alkali metals {K, Rb, Cs} and the X-site be occupied one of the halogens {Cl, Br, I}, to enable ease of synthesis and due to limited precursors availability. We then decided to computationally verify a few of the DARWIN predicted compounds (Table 1) by calculating the HSE06 bandgaps. We observe a mean absolute difference of less than 0.2 eV for ternary crystal-based systems which contain Cu as a B-site cation representing the preciseness of the framework for Cu-based systems and therefore focusing our experimental efforts.

**Table 1 | DARWIN Predicted UV Materials and their ML Bandgaps with post-DFT HSE06 Verification Calculations.**

| Predicted Structural Formula | Mean DARWIN Predicted Bandgap [eV] | HSE Bandgap [eV] | Absolute Difference [eV] | 2nd Element in Structure | Mean Absolute Difference [eV] |
|---|---|---|---|---|---|
| Cs2ZnCl4 | 3.4947 | 3.4539 | 0.0408 | Zn | 0.9592 |
| Rb2ZnCl4 | 4.0561 | 3.4123 | 0.6438 | | |
| CsZnCl3 | 3.3990 | 2.6005 | 0.7985 | | |
| SrZnBr4 | 3.1889 | 1.1973 | 1.9916 | | |
| CsZnBr3 | 3.5515 | 4.8729 | 1.3214 | | |
| Rb2CuCl3 | 3.7994 | 3.4973 | 0.3021 | Cu | 0.1930 |
| Cs2Cu1I3 | 3.5203 | 3.7664 | 0.2461 | | |
| Rb2CuBr3 | 3.6621 | 3.7054 | 0.0433 | | |
| CsCu2Br3 | 3.3244 | 3.1601 | 0.1643 | | |
| K2CuCl3 | 3.6593 | 3.5416 | 0.1177 | | |
| K2CuI3 | 3.6295 | 3.8173 | 0.1878 | | |
| Rb2CuI3 | 3.6896 | 3.4450 | 0.2446 | | |
| CsCu2Cl3 | 3.3950 | 3.6943 | 0.2993 | | |
| RbCu2Cl3 | 3.3129 | 3.4725 | 0.1596 | | |
| GaCuBr4 | 3.5037 | 3.3386 | 0.1651 | | |
| ZnMg2Cl6 | 3.8038 | 1.2318 | 2.5720 | | |
| CsMgBr3 | 3.5336 | 4.6455 | 1.1119 | | |
| CaMgBr4 | 3.8390 | 3.0983 | 0.7407 | Mg | 1.4995 |
| ZrMg2Br4 | 3.2975 | 1.2962 | 2.0013 | | |
| SrMgBr4 | 3.9983 | 2.9268 | 1.0715 | | |

As the last component of DARWIN, we pursued identifying chemical features that are responsible for structural stability and bandgap predictions. We performed an unsupervised frequency analysis of different features for the promising candidates generated by the evolutionary search. We include features beyond those included as input to the GCNN. DARWIN reconfirmed known predictive descriptors such as the A-site cations and halide choice in typical perovskite-based crystals for stability, tolerance factors, etc. (Supplementary Figures 13-16). These results showed how stability trends match the empirical tolerance factor $\tau$ - a known proxy for stable perovskite materials[33–35]. Furthermore, using Matminer's featurizers[36], several invariant feature relationships were discovered, those with a small $\sigma/\mu$ ratio ($\sigma$: standard deviation, $\mu$: mean). These constraints together ensure that the features are predictive of the properties of interest.

The top features from Figure 3a allow us to predict UV light-emitting candidates. We found that $\Delta_{X\text{-}B}$ (Figure 3b), the difference between the electronegativity (EN) of the B-site (the second most metallic element in the composition) and X-site (most electronegative anion), ranks high and exhibits a small coefficient of variation ($\sigma/\mu < 0.3$). We observe when $\Delta_{X\text{-}B}$ is within (0.95, 1.5), the material is a stable UV direct bandgap semiconductor (Supplementary Figure 13).

We denote this specific range as the Optimal Electronegativity Difference Window (OEDW).

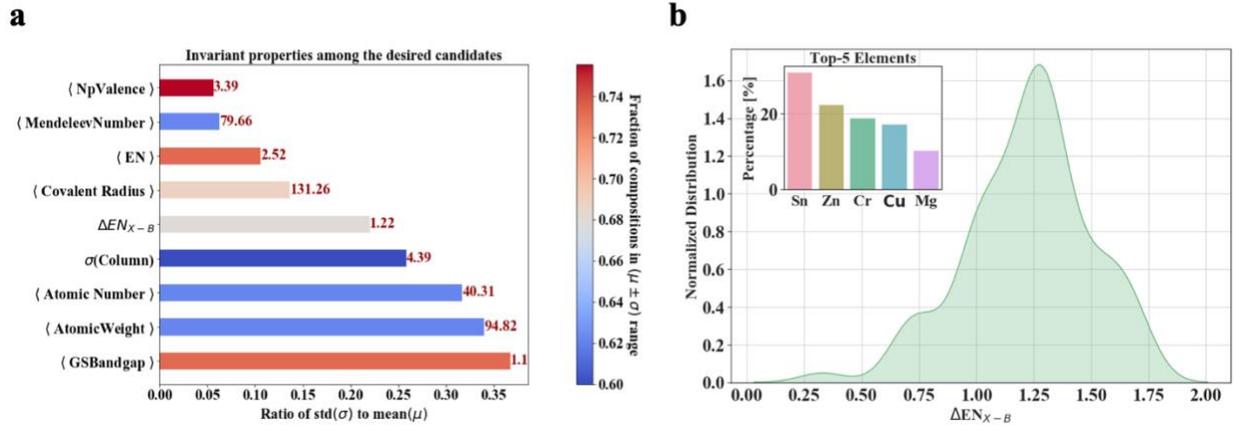

**Figure 3 | Chemical interpretability (UV Range 3.1 eV – 3.7 eV). a.** Analysis of the potential candidates generated by DARWIN by looking at the invariant properties. Color code represents the fraction present within the ($\mu \pm \sigma$) range and values at the end of each bar indicate the mean values ($\mu$). A lighter color with a small ratio indicates that the property can be used for selecting UV semiconducting candidates (GSBandgap: Gaseous state bandgap (in eV); EN: electronegativity (in eV)); Angular brackets {< and >} represent mean of the chemical quantities mentioned inside. **b.** Distribution of the difference of electronegativity between B-site (second most metallic element in a ternary) and the halide among all the successful candidates.

$K_2CuX_3$ and $Rb_2CuX_3$ (Figure 3b inset of suggested candidates) are predicted to be UV emitters with bandgaps ranging from 3.1 eV to 3.7 eV (Supplementary Figure 17 for elemental map). We performed DFT simulations to verify the predicted optical properties of $K_2CuCl_3$ (Figure 4b, refer to Supplementary Figure 18 for $K_2CuBr_3$). The results from the E-k plot (Figure 4b) indicate a direct bandgap at the Γ-point. Further analysis of the elemental contributions in the orbitally-resolved projected density of states (PDOS) reveals that the halide species significantly contributes to the valence band maxima (VBM) of such materials and the B-cation dominates the conduction band minima (CBM) (Figure 4b), thus rationalizing the observation that $\Delta_{X-B}$ is a good predictor of the bandgap. Specifically, we observe that in $K_2CuCl_3$, $K^+$ does not contribute to the electronic structure and that the strong orbital interaction of the Cu and Cl species leads to the observed optical properties[37,38].

We experimentally synthesized $K_2CuCl_3$ and $K_2CuBr_3$ (Figure 4a) via spin-coating with an intermediate anti-solvent dripping step (Methods)[39,40]. The absorption spectrum (Figure 4c) of $K_2CuCl_3$ shows an onset at 360 nm and a photoluminescence peak at 375 nm, hence a small Stokes-shift (Supplementary Figure 19). We further performed X-ray diffraction (XRD) measurements to determine the crystallized structure and observe the diffraction peaks at 14.08°, 14.70°, 15.24° and 15.86°, which match that of simulated bulk $K_2CuCl_3$ (Figure 4d, Materials Project and ICSD-150293) and throughout the 2θ range. The material crystallizes in the orthorhombic crystal system and space group *Pnma*, exhibiting 1D chains of $[CuCl_3]^{2-}$ separated by $K^+$ as shown in Figure 4a (similar to $Rb_2CuX_3$ systems)[37,38].

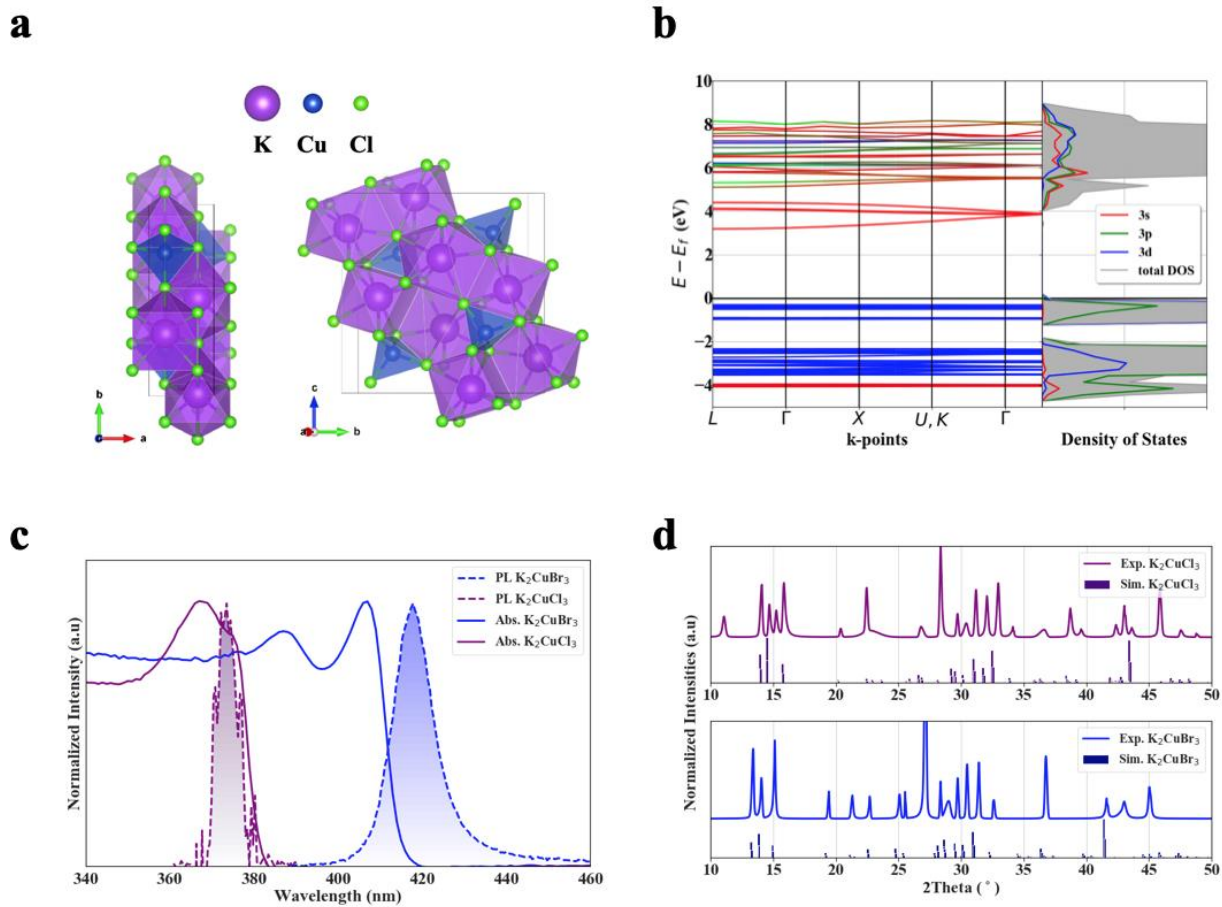

**Figure 4 | Experimental realization of K₂CuX₃ and computational studies. a.** Simulated orthorhombic crystal structure (*Pnma* space group) of $K_2CuCl_3$ illustrating the 1D chains of $[CuCl_3]^{2-}$ separated by $K^+$; **b.** Simulated band structure and Density of States of $K_2CuCl_3$ (Refer to Supplementary Information for similar analysis of $K_2CuBr_3$); **c.** The absorption spectrum and PL profiles of $K_2CuBr_3$ and $K_2CuCl_3$; **d.** Simulated[26] and experimental (powder) X-ray diffraction measurements of $K_2CuBr_3$ and $K_2CuCl_3$.

## Conclusion

DARWIN uses deep-learning models that lever electronic structure data to predict the bandgap, the structural stability estimated by the energy above hull, and the direct vs. indirect bandgap. It searches materials space using evolutionary computing coupled with deep learning.

The strategy enables efficient materials exploration and provides interpretable chemical insight. A proof-of-concept is demonstrated by focusing the search towards UV bandgap materials and successfully synthesizing candidates discovered by DARWIN.

**Acknowledgements**

This work was supported financially by the US Research Center, A Division of Sony Corporation of America (2018 Sony Research Award Program Ref# 2019-0669) and the Natural Sciences and Engineering Research Council (NSERC) of Canada. Authors thank Prof. M. Saidaminov from the University of Victoria for fruitful discussions. Computations were performed on the SOSCIP Consortium's Niagara and MIST computing platform. SOSCIP is funded by the Federal Economic Development Agency of Southern Ontario, the Province of Ontario, IBM Canada Ltd., Ontario Centres of Excellence, MITACS and 15 Ontario academic member institutions.

**Author Contributions**

E.H.S. and I.T. supervised the project. H.C., P.T. and I.T. conceived the idea. H.C, P.T. and I.T. performed the ML studies, developed the framework and methodology. J.P. and D.H.P. carried out the experimental fabrication and measurements. H.C, P.T, J.P, O.V, I.T, and E.H.S discussed the ML results. All authors discussed the results and assisted during manuscript preparation.

## METHODS.

### Data Generation for ML

For predicting stability and optoelectronic properties of the materials, we use DFT calculations to get energy above hull and bandgaps coupled with the direct/indirect nature of the band structure. We trained our modified Graph Neural Networks on total energy data obtained from the Materials Project on over 100,000 compounds and about 10,000 datapoints of direct/indirect bandgaps. The total energy values obtained from the Materials Project are based on Perdew-Burke-Ernzerhof exchange-correlation functional which has been shown to perform satisfactorily for predicting the stability of the compounds[6,20,41]. To train the bandgap regressor, we generated a small HSE06 exchange-correlation functional-based bandgap dataset. Here we performed DFT calculations using a $\Gamma$-point sampling of the Brillouin zone on ~1,800 structurally relaxed geometries obtained from the Materials Project. We invoked the flat-band approximation for the bandgap dataset. Datasets are available online as Supplementary Information.

### General Crystal Graph Network Structure.

We used the PyTorch framework and PyTorch-Geometric module to build the crystal graphs and implement the graph convolutional neural network models. The method to encode the crystal structures as graphs is a standard process previously reported in the literature[9,10,31]. Crystal structures are formatted as undirected graphs $G := \{V, E\}$ which represent nodes ($n$) as atoms and edges connecting the corresponding atoms as bonds, respectively. Each node and edge can have a corresponding feature vector, which fully describes the 2D planar representation of the crystal structure.

In our network, we encode atomic features denoted $u_i$ (a particular set of physical and chemical features native to the atom at node $i$, refer to Supplementary Information for an exhaustive list) and a new edge feature vector labelled $e_{i,j}$ as the reciprocal distance between the connecting nearest neighbouring atoms $i$ and $j$. Nearest neighbours at each node are selected by implementing a cut-off radius of 8Å, ensuring that only the local environment is considered.

The resulting crystal graph allows for a convolutional neural network to be built on top. The general idea of a graph convolutional neural network is to take a graph as input; perform a particular convolution operation (consisting of a message and update step) based on the nodal and edge-based feature vectors, and output a new scalar or vector for each corresponding node and edge. This convolution operation is repeated $T$ iterations, at which point the newly updated features are then pooled together to obtain an aggregated prediction value by a read-out function. At each successive iteration, the nodal features are updated to incorporate long-range order from atoms outside of the initial cut-off radii; learning a feature vector that includes information from the surrounding environment and enabling accurate predictions of material properties inherent to the crystal structure. The convolution operator for an atom feature vector $u_i$, in this case,

$$u_i^{t+1} = Conv(u_i^t, u_j^t, e_{ij}), \forall\, i,j \in G,$$

where the indices $i$ and $j$ correspond to two atoms connected in graph $G$. The pooling operation to aggregate all of the newly learnt hidden nodal features to predict a value for a given crystal is shown below and is typically defined by a function such as the mean, maximum or sum:

$$\hat{y} = Pool(u^T), \forall\, u \in G$$

All code and models are available online as Supplementary Information.

**Predictive Graph-based Convolutional Neural Network Models.**

*Bandgap Models*

Many studies have implemented various convolutional operators with success[6,10,31,42], but here we use a complimentary set of convolutional graph neural networks as the basis for our predictive ML models. We design a specific convolution operation for the bandgap regressor shown below. Each crystal graph is fed into the network as an input; wherein the convolutional layer is defined by the following order of operations: the maximum vector after concatenation ($\oplus$) of the current atoms feature vector ($u_i^t$), the product of each neighbouring ($\mathcal{N}(i)$) atomic feature vectors ($u_j^t$) with the corresponding edge features ($e_{ij}$) at each iteration $t$. The $\gamma^t$ represents the update function (typically defined by a multilayer perceptron (MLP) consisting of a non-linear activation function $g$) and weight tensors $W_s^t$ and $b^t$ which are learnt during the training steps for each convolutional layer.

$$u_i^{t+1} = \gamma^t \left[ \max_{j \in \mathcal{N}(i)} \left( u_i^t \oplus (u_j^t \cdot e_{ij}) \right) \right], \quad \begin{array}{l} \forall\, i,j \in G \\ \forall\, t \in \{1, \ldots, T\} \end{array}$$

$$u_i^{t+1} = g \left[ \max_{j \in \mathcal{N}(i)} \left( u_i^t \oplus (u_j^t \cdot e_{ij}) \right) W_s^t + b^t \right], \quad \begin{array}{l} \forall\, i,j \in G \\ \forall\, t \in \{1, \ldots, T\} \end{array}$$

An output value is obtained by pooling all of the hidden nodal features using the mean pooling function which results in the prediction of a bandgap value $\hat{y}_{bandgap}$.

$$\hat{y}_{bandgap} = \frac{1}{n} \sum_{i=1}^{n} u_i^T, \quad \forall\, u \in G$$

To measure the accuracy of the ML models, we use the mean squared error as a cost/loss function. We seek to minimize the MSE during training which evaluates how well our predicted bandgap is with respect to the target values. We achieve this minimization by finding the optimal values for the weight and bias matrices in the corresponding convolutional layers.

$$MSE = \frac{1}{n} \sum_{i=1}^{n} (y_i - \hat{y}_i)^2, \quad \forall\, i \in G$$

We observed that a total of 4 convolutional layers and 2 fully connected dense layers were the most optimal choice of the network that led to accurate predictions of the bandgaps. The list of atomic features (Supplementary Tables 1 and 2) and the best performing set are reported in the Supplementary Information. All features were normalized via standard ML procedures prior to the crystal graph generation to ensure comparable scalar magnitudes. Graphs were generated using the aforementioned process, enabling efficient and rapid batch training. Hyperparameters of the model were optimized which include: the size of each convolutional layer, choice of activation function, pooling layer, learning rate and weight decay. After hyperparameter optimization, we found the following parameters: Adam optimizer, a rectifying linear unit activation function, learning rate of 0.01, weight-decay of 0.0005, 4 convolutional layers (consisting of 64, 64, 32, 16 output channels) resulted in a mean-absolute-error of 0.46 eV on validation data after 3000 epochs of training on a set of roughly 1,800 crystals. The best model was chosen and used as one of the surrogate models for the evolutionary algorithmic approach.

*Energy Models*

Using the same convolution expressions as the bandgap model, 4 convolutional layers with batch normalization and 2 fully connected dense layers were the best network that led to accurate predictions of the energy above hull. After hyperparameter optimization that the following parameters: Adam optimizer, a rectifying linear unit activation function, learning rate of 0.0001, 4 convolutional layers (consisting of 64, 64, 64, 64 output channels) resulted in a mean-absolute-error of 0.06 eV/atom on validation data after 2000 epochs of training on a set of roughly 100,000 crystals (refer to SI for loss curves).

*Direct-Indirect classification*

For the direct-indirect bandgap classification, we use a random forest classifier trained on global features generated using elemental compositions and Spacegroup information.

Using statistical properties of elemental properties such as electronegativity and atomic mass, we generate a global representation using Matminer. TPOT was then employed to find the best hyper-parameters. The resulting ML pipeline achieved an $F_1$-score of 0.8 on direct-indirect classification. (refer to SI table 3 for the list of the properties used, and SI note 4 for ML model parameters)

**Evolutionary Algorithm.**

The EA operates on a surrogate model composed of the three predictive ML models built for the various prediction and classification tasks. A selection criterion is designed for target material properties such as the bandgap value and stability. In general, the multi-step iterative process by which the evolutionary search is implemented is as follows: (1) initialization of primary candidates denoted as the initial generation; (2) prediction of material properties using the ML models; (3) evaluation of the current generation; (4) selection of the fittest candidates; and (5) mutations in the selected individuals, and developing a new generation of candidates. Over successive iterations, the evolutionary algorithm converges and outputs a set of candidates that are optimal given the current set of selection parameters. *Initialization*: In the initialization step, we select a set of elements and generate an initial set of candidates based on the 200 crystal structure types and 7 families. We select the bandgap and energy above hull which we would like to optimize for and set these search criteria. *Prediction*: Crystal graphs are generated via the aforementioned process and fed as inputs into the three pre-trained ML models to obtain prediction values for the bandgap, energy above hull, and direct-indirect classification. *Evaluation:* We evaluate each individual in the current generation given the loss metric as shown in the equation below which is a weighted sum of the squared loss for each individually predicted property and the target selection values, where $\lambda_i$ are normalizing factors for each loss component. For the selection procedures, we set all the weights to be equal. We initialize with a population of 20 arbitrarily formed prototype structures, set the generations limit threshold at 200. *Selection*: Upon evaluating the loss, we rank all individuals by their loss in the current generation and discard the bottom-half and retain the remaining population. *Mutation*: We then proceed to make a mutation on each top-ranked individual in the population which we define as a single elemental substitution in the crystal structure with the equivalent oxidation state to retain structural charge-neutrality. The new set of candidates is then added to the current top-ranked generating a new population and the process is repeated but now starting at evaluation. After multiple iterations the loss has plateaued, the EA proposes a set of candidate solutions that ideally match the initial selection criteria. The proposed crystal structures are then aggregated and collected to comprise of the candidate solutions for the given target conditions. This process is repeated (100 times in our experiments) for various selection criteria to span the varied bandgap range and design a set of candidate solutions for further analysis and experimental realization.

$$\mathcal{L} = \lambda_1 \left(\hat{E}_{gap} - E_{gap}^{target}\right)^2 + \lambda_2 \left(\hat{E}_{hull} - E_{hull}^{target}\right)^2 + \lambda_3 \left(\hat{E}_{direct} - 1\right)^2$$

**Experimental Synthesis – Film Fabrication.**

Potassium halide (KX, X = I, Br, Cl), copper halide (CuX, X = I, Br, Cl), dimethylsulfoxide (DMSO) and dimethylformamide (DMF) were purchased from Sigma-Aldrich. Chloroform was purchased from DriSolv. All chemicals were used as received. The precursor solution was prepared by dissolving stoichiometric quantities of KX and CuX in a DMSO/DMF (25/75 % v/v) solution (0.5 M) under continuous stirring for 1 h at room temperature. The concentration of the chloride-based precursor solution (in DMSO/DMF 75/25 % v/v) was limited to 0.2 M due to the low-solubility of the precursors. Glass substrates were $O_2$ plasma-treated to improve adhesion. The precursor solution was spin-coated onto the substrates via a two-step process: 1000 rpm for 10 s and 3000 rpm. for 60 s. During the second spin step, 0.5 mL of chloroform was poured onto the substrate. The films were then annealed at 110 °C for 10 min. All the samples were prepared in a glove box with $N_2$ atmosphere in order to control the atmospheric conditions.

**Material characterization.**

X-ray diffractograms were recorded using a Rigaku MiniFlex 600 powder X-ray diffractometer equipped with a NaI scintillation counter and using monochromatized Cu Kα radiation (l=1.5406Å). UV−Vis absorption was measured using a Perkin Elmer LAMBDA 950 UV/Vis/NIR spectrometer. PL measurements were collected using a UV-Vis USB 2000+ spectrometer (Ocean Optics). The samples were optically excited using a 355 nm frequency-tripled Nd:YAG laser with a pulse width of 2 ns and a repetition rate of 100 Hz.

**Code Availability.**

All the necessary codes and generated HSE06 data can be found in the attached zip file. Open source data was obtained from Materials Project procured using Pymatgen MPRester API.